\documentclass[conference]{IEEEtran}

\usepackage{cite}
\usepackage{graphicx}

\usepackage[top=0.7in,bottom=1.049in,left=0.68in,right=0.68in]{geometry}
\usepackage[cmex10]{amsmath}
\usepackage{url}
\usepackage{amssymb}
\usepackage{mathtools}
\usepackage{bm}
\usepackage{algorithm,algpseudocode}
\usepackage{booktabs}
\usepackage{multirow}
\usepackage{subcaption}

\usepackage{comment}

\def\eg{\textit{e.g.}}

\DeclareMathOperator*{\argmax}{arg\,max}

\def\BibTeX{{\rmfamily B\kern-.05em{\scshape i\kern-.025em b}\kern-.08em
 T\kern-.1667em\lower.7ex\hbox{E}\kern-.125em X}}

\makeatletter
 \def\subsubsection{\@startsection{subsubsection}
                                 {3}
                                 {\z@}
                                 {0ex plus 0.1ex minus 0.1ex}
                                 {0ex}
                                 {\normalfont\normalsize\itshape}}
\makeatother

\begin{document}
\title{
MAB-based Client Selection for Federated Learning with Uncertain Resources in Mobile Networks
}

\author{
\IEEEauthorblockN{
Naoya Yoshida\IEEEauthorrefmark{1},
Takayuki Nishio\IEEEauthorrefmark{1},
Masahiro Morikura\IEEEauthorrefmark{1}, and
Koji Yamamoto\IEEEauthorrefmark{1}
}

\IEEEauthorblockA{\IEEEauthorrefmark{1}
Graduate School of Informatics, Kyoto University, Yoshida-honmachi, Sakyo-ku, Kyoto 606-8501, Japan
}
\IEEEauthorblockA{\IEEEauthorrefmark{1}
nishio@i.kyoto-u.ac.jp
}
}

\maketitle
\begin{abstract}
  This paper proposes a client selection method for federated learning (FL) when the computation and communication resource of clients cannot be estimated; the method trains a machine learning (ML) model using the rich data and computational resources of mobile clients without collecting their data in central systems.
  Conventional FL with client selection estimates the required time for an FL round from a given clients' computation power and throughput and determines a client set to reduce time consumption in FL rounds.
  However, it is difficult to obtain accurate resource information for all clients before the FL process is conducted because the available computation and communication resources change easily based on background computation tasks, background traffic, bottleneck links, etc.
  Consequently, the FL operator must select clients through exploration and exploitation processes.
  This paper proposes a multi-armed bandit (MAB)-based client selection method to solve the exploration and exploitation trade-off and reduce the time consumption for FL in mobile networks. The proposed method balances the selection of clients for which the amount of resources is uncertain and those known to have a large amount of resources.
  The simulation evaluation demonstrated that the proposed scheme requires less learning time than the conventional method in the resource fluctuating scenario.
\end{abstract}
\IEEEpeerreviewmaketitle

\section{Introduction}\label{sec:Intro}
Collaborative machine learning (ML), which leverages the data and computational resources of mobile devices to train ML models, has attracted substantial attention.
Currently, there are nearly 7 billion connected Internet of Things (IoT) devices worldwide \cite{NumOfIotDevices} and 3 billion smartphones.
These devices have increasingly advanced sensing, computing, and communication capabilities, and these resources can be used to empower modern artificial intelligence (AI) products via cutting-edge ML techniques.
Federated learning (FL), which is a framework of collaborative ML, trains an ML model using the rich data and computational resources of the mobile clients without collecting their data in central systems \cite{DBLP:conf/aistats/McMahanMRHA17}.
In FL, an edge server or a cloud server assigns ML tasks to mobile clients and gathers the models updated by the mobile clients using their own data, instead of gathering data and training models by itself.
Because the data is kept in the local storage of mobile clients, private data can be used to train the model.

One bottleneck for FL is the latency caused by wireless transmission and local computation on clients, especially when the number of participating clients in the FL process is large \cite{DBLP:journals/corr/abs-1712-01887,abs-1902-01046}.
This is because mobile clients with differing wireless-link qualities and computational capabilities must update and communicate a several-gigabyte ML model.
One important approach to reducing the training latency in FL is client selection.
The FL operator must determine how much bandwidth should be allocated to each client in order to determine which clients should participate in FL.
Many client selection methods in FL have been proposed in existing studies.
In \cite{nishio2018client, yoshida2020}, client selection algorithms that maximize the number of selected clients in each round were studied to increase accuracy improvement per round.
The authors in \cite{9053740} proposed a client scheduling algorithm to jointly consider the staleness of the received parameters and the instantaneous channel qualities.
A joint scheduling of wireless resources and client scheduling was studied in \cite{chen2019joint}.
These studies assume that accurate resource information (e.g., throughput, channel state, and available computation power of clients) is given, and it is assumed that the amount of resources is stable in a round.
However, to obtain the resource information before conducting FL processes, resource measurements are required (e.g., beaconing, throughput prediction, and computation load estimation), which consume computation and communication resources.
Moreover, the obtained resource information may not always be accurate and stable in the FL processes due to background computation tasks, background traffic, unstable wireless link quality, and network congestion.
Therefore, the existing client selection approach based on the ideal assumption for resource information could suffer from large performance degradation induced by mismatch between the mathematically obtained optimal selection and the actual optimal selection in practice.

This paper proposes a trial-and-error based client selection algorithm based on the multi-armed bandit (MAB) algorithm for use in situations in which the client resources are not exactly determined.
We achieve more efficient scheduling by performing exploration, which selects clients that are selected less frequently, and exploitation, which selects clients with rich resources, and balancing them effectively.
The MAB problem is a sequential decision problem, where the player selects an arm in each timeslot and observes the reward (payoff).
The aim of the MAB problem is to determine the arm so as to maximize the total reward obtained in sequential selections.
The application of MAB to resource scheduling in wireless networks has been widely studied \cite{9023919,8895788}.
Our main contribution is providing a FL protocol that incorporates MAB as a method for solving the client selection problem in realistic environments wherein the client resources are uncertain and fluctuate.
The proposed client selection algorithm finds and leverages resource-rich clients by trial and error based on MAB to improve the trade-off between exploration and exploitation in the FL.
We evaluate our method using realistic large-scale training simulations involving neural networks for image classification, with a cellular network as the simulation environment.
The simulation results demonstrate that the proposed protocol achieves higher classification accuracy than the existing protocol when the client resources are highly variable.

\section{System Model}\label{sec:SystemModel}
The system model follows a model described in a previous work \cite{nishio2018client}.
We consider a certain platform, which is located in a cellular network and consists of an edge server and a base station (BS).
The FL server manages the behaviors of the server and clients in the FL protocol.

We assume that there is a fixed number of resource blocks (RBs) \cite{sesia2011lte} available in the cell and that they are shared among the clients.
In addition, if multiple clients communicate with the server simultaneously, the throughput for each client decreases accordingly.
We also assume that the modulation and coding schemes for radio communication are determined suitably for each client, while considering its channel state and packet-loss rate to be negligible.
This leads to different throughput for each client communicating with the server, although the number of allocated RBs is constant.

This work considers that the client resource information consists of two values: the time needed for the client to update the model, and the time needed to upload the model to the FL server.
In addition, we assume that all clients' model update times and throughputs fluctuate based on a certain distribution that is unknown to the FL operator.
Some clients consume communication and computational resources owing to tasks unrelated to FL.
Therefore, it is natural to assume that the resources are variable.
A client's model update time and model upload time are set to the actual times consumed by participating in FL.
When a client is selected and participates in FL, the resource information is updated with the new values.
The client notifies the server of the values until the client is selected again.
Note that clients participating in FL for the first time notify the server that their model update time and model upload time are $0$\,s.

FL protocol is shown as follows.
In FL, the server first randomly initializes a global model, after which the following steps are executed iteratively.
In the \texttt{Resource Request} step, $\lceil K\times C \rceil$ random clients (where $K$ denotes the total number of clients, $C\in(0, 1]$ denotes a hyperparameter representing the proportion of clients participating in each round to the total number of clients, and $\lceil\cdot\rceil$ denotes the ceiling function) inform the FL server about their resource information, such as wireless-channel states, computational capacities, and data amounts relevant to the current training task (\eg, if the server intends to train a ``dog-vs-cat'' classifier, then the number of images containing dogs or cats will be specified).
In the \texttt{Client Selection} step, the FL server uses the above information to determine which of the clients will proceed to the subsequent steps.
In the \texttt{Distribution} step, the server then distributes the parameters of the global model to the selected clients.
In the \texttt{Model Update} step, the selected clients update the global models in parallel by using their own data; they then upload the new parameters to the server by using the RBs allocated by the FL server.
Finally, in the \texttt{Aggregation} step, the server aggregates multiple models updated by the selected clients in order to improve the global model.
All steps except \texttt{Initialization} are iterated multiple times until the global model achieves the desired performance or the final deadline arrives.

\section{Multi-Armed Bandit Algorithm-Based Client Selection}\label{sec:proposed}
We propose a new client selection algorithm for use in the \texttt{Client Selection} step of the FL protocol explained in Sect.~\ref{sec:SystemModel}, which performs efficiently in situations in which client resources fluctuate and are unknown until a FL round is conducted.
The proposed algorithm uses the MAB algorithm to estimate which clients are expected to have rich and available computation power and throughput.
In the proposed FL, the server receives the model update and model upload times consumed by the selected clients as the reward in the \texttt{Scheduled Upload} step.

The key idea behind our algorithm is that we adapt the MAB algorithm to achieve efficient scheduling by performing exploration, which selects clients that are selected less often, and exploitation, which selects clients with rich resources, and balancing them effectively.
Our algorithm is based on the upper confidence bound (UCB) policy \cite{bandit}, which is the most elementary MAB algorithms.

\subsection{Client Selection in FedCS}
In this section, we introduce the client selection algorithm in FedCS \cite{nishio2018client}.
In FedCS, a set of clients are selected as shown in Algorithm~\ref{alg:scheduling}.
Here,
$T_{\mathrm{inc}}(\bm{S},k)$ denotes the estimation time, which signifies how much the round time will increase when adding client $k$ to $\bm{S}$, and
$f(\bm{S},k)$ denotes the client evaluation value.
In the FedCS protocol, $f(\bm{S},k)$ is $-T_{\mathrm{inc}}(\bm{S},k)$.
We iteratively add the client that consumes the least model upload time, and subsequently update $\bm{S}$ until the number of selected clients $|\bm{S}|$ reaches the required number $S_{\mathrm{round}}$.

\begin{figure}[t]
  \vspace{-3mm}
  \makebox[\linewidth]{
  \begin{minipage}{\linewidth}
  \begin{algorithm}[H]
  \floatname{algorithm}{Algorithm}
  \caption{Client selection}
  \label{alg:scheduling}
  \begin{algorithmic}[1]
  \Require{Index set of randomly selected clients $\bm{K}'$}
  \State \textbf{Initialization} $\bm{S} \gets \{\}$, $t\gets 0$ 
  \While{$|\bm{K}'| > 0$}
    \State{$x \gets \argmax_{k\in \bm{K}'} f(\bm{S},k)$}
    \State{remove $x$ from $\bm{K}'$}
    \State{$t^{\prime} \gets t+T_{\mathrm{inc}}(\bm{S},x)$}
    \If{$|\bm{S}|<S_{\mathrm{round}}$}
      \State{$t \gets t^{\prime}$}
      \State{add $x$ to $\bm{S}$}
    \EndIf{}
  \EndWhile{}\\
  \Return{$\bm{S}$}
  \end{algorithmic}
  \end{algorithm}
  \end{minipage}
  }
\end{figure}

The details for estimating $T_{\mathrm{inc}}(\bm{S},k)$ are shown as follows.
$T_{\mathrm{inc}}(\bm{S},k)$ is calculated based on the model update time and model upload time as follows.
Let $\bm{K} = \{1,\dots,K\}$ be a set of indices that describes $K$ clients and let $\bm{K}^\prime \subseteq \bm{K}$ be a subset of $\bm{K}$ randomly selected in the \texttt{Resource Request} step.
$\bm{S} = [k_1,\dots,k_i,\dots,k_{|\bm{S}|}]$, where $k_i\in \bm{K}^\prime$, $|\bm{S}|\leq |\bm{K}^\prime|$, denotes a sequence of indices of the clients selected in the \texttt{Client Selection} step.
Let $t_k^{\mathrm{UL}}$ be the model upload time of client $k$, $t_k^{\mathrm{UD}}$ be the model update time of client $k$, and $T_{\bm{S}}^{\mathrm{d}} = \max_{k_i\in\bm{S}}\{t_{k_i}^{\mathrm{UL}}\}$ be the time required for the \texttt{Distribution} step; $t_k^{\mathrm{UL}}$, $t_k^{\mathrm{UD}}$, and $T_{\bm{S}}^{\mathrm{d}}$ are non-negative real numbers.
$T_{\mathrm{inc}}(\bm{S},k)$ is given by
\begin{align}
  T_{\mathrm{inc}}(\bm{S},k) &= \left(T_{\bm{S}\cup k}^{\mathrm{d}}-T_{\bm{S}}^{\mathrm{d}}\right)+\max\{t_k^{\mathrm{UD}}-(t-T_{\bm{S}}^{\mathrm{d}}),0\}+t_k^{\mathrm{UL}}. \label{eq:T_inc}\\
  T_{\bm{S}}^{\mathrm{d}} &= \max_{k_i\in\bm{S}}{t_{k_i}^{\mathrm{UL}}}
\end{align}

As mentioned in Sect.~\ref{sec:Intro}, the performance of FL combined with FedCS will degrade when client resources fluctuate.
After $T_{\mathrm{inc}}(\bm{S},k)$ is calculated based on the resource information obtained from the clients in the \texttt{Resource Request} step, the actual time required for the \texttt{Model Update} and \texttt{Scheduled Upload} steps will be different from $T_{\mathrm{inc}}(\bm{S},k)$ because of the resource fluctuation.
Thus, some clients consume more time than $T_{\mathrm{inc}}(\bm{S},k)$ and the learning time of FL may increase.

\subsection{Straightforward Application of MAB to Client Selection}
Here, we consider naively applying MAB to the client selection problem.
The typical application of MAB is slot machine selection, in which a player selects an arm of slot machine from multiple slot machines.
The player selects and plays an arm, after which they obtain a reward.
By repeating this trial-and-error process, the player learns a strategy to optimally select an arm to maximize the expected reward.
In MAB, the selection problem is formulated as a problem of selecting an arm from among multiple arms so that the cumulative reward is maximized.
In the UCB algorithm, the player selects an arm based on the following UCB score:
\begin{align}
  \overline{\mu_k} + \sqrt{\frac{\log{(\Sigma N_k)}}{2N_k}},
\end{align}
where $\overline{\mu_k}$ denotes the sample mean of the k-th arm reward $\mu_k\in[0,1]$, and $N_k$ denotes the number of times that k-th arm has been selected.

In the client selection, the FL operator selects a set of clients participating the FL round.
By applying the MAB naively to the client selection, the arms of the MAB correspond to sets of clients.
The MAB algorithm struggles to converge because the number of arms increases combinatorially as the number of clients increases.
Specifically, in FL, the number of arms reaches on the order of ten to the tenth power or more, since about ten clients are selected in each round from among approximately a thousand clients.

\subsection{Proposed MAB-based Client Selection}
We designed the client selection in the framework of the MAB problem so that the number of arms is not excessively large.
The proposed algorithm creates a set of clients by adding each client sequentially, as with FedCS.
We consider each client added to a set as an arm of the MAB framework, which allows us to reduce the number of arms to be the same as the number of clients.
The FL server calculates the UCB score for each $\lceil K\times C \rceil$ client and selects the $S_{\mathrm{round}}$ clients in order of score.
The proposed algorithm is expressed in the same form as in Algorithm~\ref{alg:scheduling}, where the evaluation value $f(\bm{S},k)$ is set to the UCB score of each client.
We propose two designs for UCB scores as follows.

\subsubsection{Naive UCB Score}
First, we set the payoff to the increased time $T_{\mathrm{inc}}(\bm{S},k)$ and evaluate each client by the UCB score as follows.
Let $\overline{T_{\mathrm{inc}}}(\bm{S},x)$ be the sample mean of time denoting how much the round time has been extended when adding client $k$ to $\bm{S}$.
$\overline{T_{\mathrm{inc}}}(\bm{S},x)$ is updated using the reward obtained by the FL server in the \texttt{Scheduled Upload} step.
$f(\bm{S},k)$, which is the evaluation value of client $k$, is set to the Naive UCB Score given by
\begin{align}
  f(\bm{S},k) = -\overline{T_{\mathrm{inc}}}(\bm{S},k)/\alpha + \sqrt{\frac{\log{(\Sigma N_k)}}{2N_k}},\label{eq:evaluation1}
\end{align}
where $N_k$ denotes the number of times that client $k$ has been selected.
The value of constant weight $\alpha$ is similar to that of $T_{\mathrm{inc}}(\bm{S},k)$.
Since $T_{\mathrm{inc}}(\bm{S},k)$ is not in $[0,1]$, $\alpha$ simply balances the first and second term.
The right-hand side is the UCB score.
The less frequently a client is selected, the smaller the amendment term of the client will be.
Therefore, clients with a small number of selections are more likely to be selected, even if the client's round times are long.
This algorithm aims to directly reduce the round time by utilizing the reward setting.
However, convergence is not guaranteed by this UCB adaptation because the reward depends on the other clients' payoff.

\subsubsection{Element-wise UCB Score}
Second, we design the payoffs such that they are not affected by the other arms, as assumed in the UCB algorithm.
Therefore, the payoff is set to the model update time and model upload time, which are determined independently for each client.
$T_{\mathrm{inc}}(\bm{S},k)$ depends only on the model update time and model upload time.
To adapt the UCB algorithm with these payoffs to the client selection, we introduce the following values.
\begin{align}
  \tau_k^{\mathrm{UD}} &= \overline{t_k^{\mathrm{UD}}}/\beta-\sqrt{\frac{\log{(\Sigma N_k)}}{2N_k}},\label{eq:tau_UD}\\
  \tau_k^{\mathrm{UL}} &= \overline{t_k^{\mathrm{UL}}}/\beta-\sqrt{\frac{\log{(\Sigma N_k)}}{2N_k}},\label{eq:tau_UL}
\end{align}
where $\overline{t_k^{\mathrm{UL}}}$ denotes the sample mean of the model upload time of client $k$, and $\overline{ t_k^{\mathrm{UD}}}$ denotes the sample mean of the model update times of client $k$.
The constant weight $\beta$ is set for the same purpose as $\alpha$.
$\overline{t_k^{\mathrm{UL}}}$ and $\overline{t_k^{\mathrm{UD}}}$ are updated using the reward obtained by the FL server in the \texttt{Scheduled Upload} step.
$\tau_k^{\mathrm{UD}}$ and $\tau_k^{\mathrm{UL}}$ denote the model update time and model upload time with negative UCB amendment terms.
The second terms in Eq.~\eqref{eq:tau_UD} and Eq. \eqref{eq:tau_UL} are negative UCB amendment terms. The fewer times a client $k$ is selected, the smaller $\tau_k^{\mathrm{UD}}$ and $\tau_k^{\mathrm{UL}}$ will be.
$f(\bm{S},k)$ is set to the Element-wise UCB Score calculated using $\tau_k^{\mathrm{UD}}$ and $\tau_k^{\mathrm{UL}}$ as follows.
\begin{align}
  f(\bm{S},k) &= -T'_{\mathrm{inc}}(\bm{S},k),
\end{align}
where $T'_{\mathrm{inc}}(\bm{S},k)$ is given by replacing $t_k^{\mathrm{UD}}$ and $t_k^{\mathrm{UL}}$ with $\tau_k^{\mathrm{UD}}$ and $\tau_k^{\mathrm{UL}}$ in Eq.~\eqref{eq:T_inc}, respectively.
Therefore, clients with a small number of selections are more likely to be selected, even if the client's model upload time and model update time are relatively long.
In this algorithm, the payoffs are determined independently for each client;
consequently, it is expected that resource-rich clients can be selected more efficiently.

\section{Performance Evaluation}\label{sec:Evaluation}
As a proof of concept that our protocol works effectively, we conducted simulations by performing realistic ML tasks using publicly available large-scale datasets.
The simulation is inspired from \cite{nishio2018client}.
We evaluated the performance of our protocol under varying resource fluctuation.

\subsection{Simulation Settings} \label{subsec:SimulationSettings}
The simulation environment comprised an edge server, a BS, and $K=100$ clients.
The BS and server were co-located at the center of a 2-km radius cell, and the clients were uniformly distributed in the cell.
The computational capability of the server was sufficiently high compared to that of the clients.
Therefore, the time required for \texttt{Client Selection} could be ignored.
We set hyperparameter $C=0.1$ based on \cite{DBLP:conf/aistats/McMahanMRHA17} and set the number of clients selected in a round $S_{\mathrm{round}} = 5$.
We also set the parameters $\alpha = 1000$, similar to that of $T_{\mathrm{inc}}$, and $\beta = 50$, similar to that of $t_k^{\mathrm{UL}}$ and $t_k^{\mathrm{UD}}$ in a no-resource-fluctuation setting.

Wireless communications were modeled based on long-term evolution (LTE) networks with an urban channel model defined in the ITU-R M.2135-1 Micro NLOS model with a hexagonal cell layout~\cite{ITU-R}.
Further, we set the model parameters as follows.
The carrier frequency was 2.5\,GHz; the antenna heights of the BS and clients were 11 and 1\,m respectively; and the transmission power and antenna gain of the BS and clients were 20 and 0\,dBi, respectively.
As a practical bandwidth limitation, we assumed that 10 RBs, which corresponded to a bandwidth of 1.8 MHz, were assigned to a client in each time slot of 0.5\,ms.
The throughput model was based on the Shannon capacity with a certain loss used in~\cite{6834753} with $\Delta = 1.6$ and $\rho_\mathrm{max}=4.8$.
The mean and maximum throughputs of the clients were 1.4 and 8.6\,Mbit/s, respectively, both of which are realistic values in an LTE network.
We considered the throughput $\theta_k$ obtained from the above-mentioned model as the average throughput of each client.
The throughput in the simulation was sampled from a truncated normal distribution after the clients send their resource information to the server.
The distribution was determined as follows.
Let $\mathbb{R}$ be a set of real numbers ($\mu, \sigma, a, b\in\mathbb{R}$), and $a\leq\mu\leq b$.
A cumulative distribution function of the truncated normal distribution for $a\leq x\leq b$ is given by
\begin{align}
    F(x ; \mu, \sigma, a, b)=\frac
    {\Phi\Bigl(\frac{x-\mu}{\sigma}\Bigr)-\Phi\Bigl(\frac{a-\mu}{\sigma}\Bigr)}
    {\Phi\Bigl(\frac{b-\mu}{\sigma}\Bigr)-\Phi\Bigl(\frac{a-\mu}{\sigma}\Bigr)}\label{eq:TND}.
\end{align}
Further, $\Phi(\cdot)$, which denotes the cumulative distribution function of the standard normal distribution, is given by
\begin{align}
  \Phi(x)=\frac{1}{2}\left(1+\operatorname{erf}\!\left(x / \sqrt{2}\right)\right).
\end{align}
Let $\eta \in \mathbb{R}, \eta < 2$ be a parameter that denotes the amount of resource fluctuation.
In the simulation, we sampled the throughput from the truncated normal distribution in Eq.~\eqref{eq:TND}, where $\mu = \theta_k, \sigma^2 = \theta_k^\eta, a=\theta_k-\sigma, b=\theta_k+\sigma$.
Thus, the larger the value of $\eta$, the greater the amount of resource fluctuation.

The computational capability was modeled based on how many data samples could be processed per second to update a model;
moreover, computational capability can fluctuate because of other computational loads.
Subsequently, we randomly determined the average computational capability of each client, $\gamma_k$, within a range of 10 to 100.
Similar to the calculation for communication capability, the computational capability in the simulation was sampled from a truncated normal distribution in Eq.~\eqref{eq:TND}, where $\mu = \gamma_k, \sigma^2 = \gamma_k^\eta, a=\gamma_k-\sigma, b=\gamma_k+\sigma$, at the same time as the throughput sampling.

The model update time $t_k^{\mathrm{UD}}$ and model upload time $t_k^{\mathrm{UD}}$ of client $k$ can be determined as follows from the sampled throughput $\theta_k^{\mathrm{tmp}}$ and computational capability $\gamma_k^{\mathrm{tmp}}$.
\begin{align}
  t_k^{\mathrm{UD}} &= D_k/\gamma_k^{\mathrm{tmp}},\\
  t_k^{\mathrm{UL}} &= M/\theta_k^{\mathrm{tmp}},
\end{align}
where $D_k$ denotes the number of data samples stored in client $k$ and $M$ denotes the data size of the ML model.
Each client sends $t_k^{\mathrm{UD}}$ and $t_k^{\mathrm{UL}}$ to the server in the \texttt{Resource Request} step.

\subsection{ML Setups}
We adopted a realistic object-classification task using large-scale image datasets in the simulated environment.
The dataset was CIFAR-10 \cite{krizhevsky2009learning}, which is a classic object-classification dataset comprising 50,000 training images and 10,000 testing images, with 10 object classes.
This dataset has been employed in numerous other FL studies \cite{DBLP:conf/aistats/McMahanMRHA17,Konecny2016}.

\begin{figure}[t]
  \centering
  \subcaptionbox{No resource fluctuations}{
  \includegraphics[width=0.9\columnwidth]{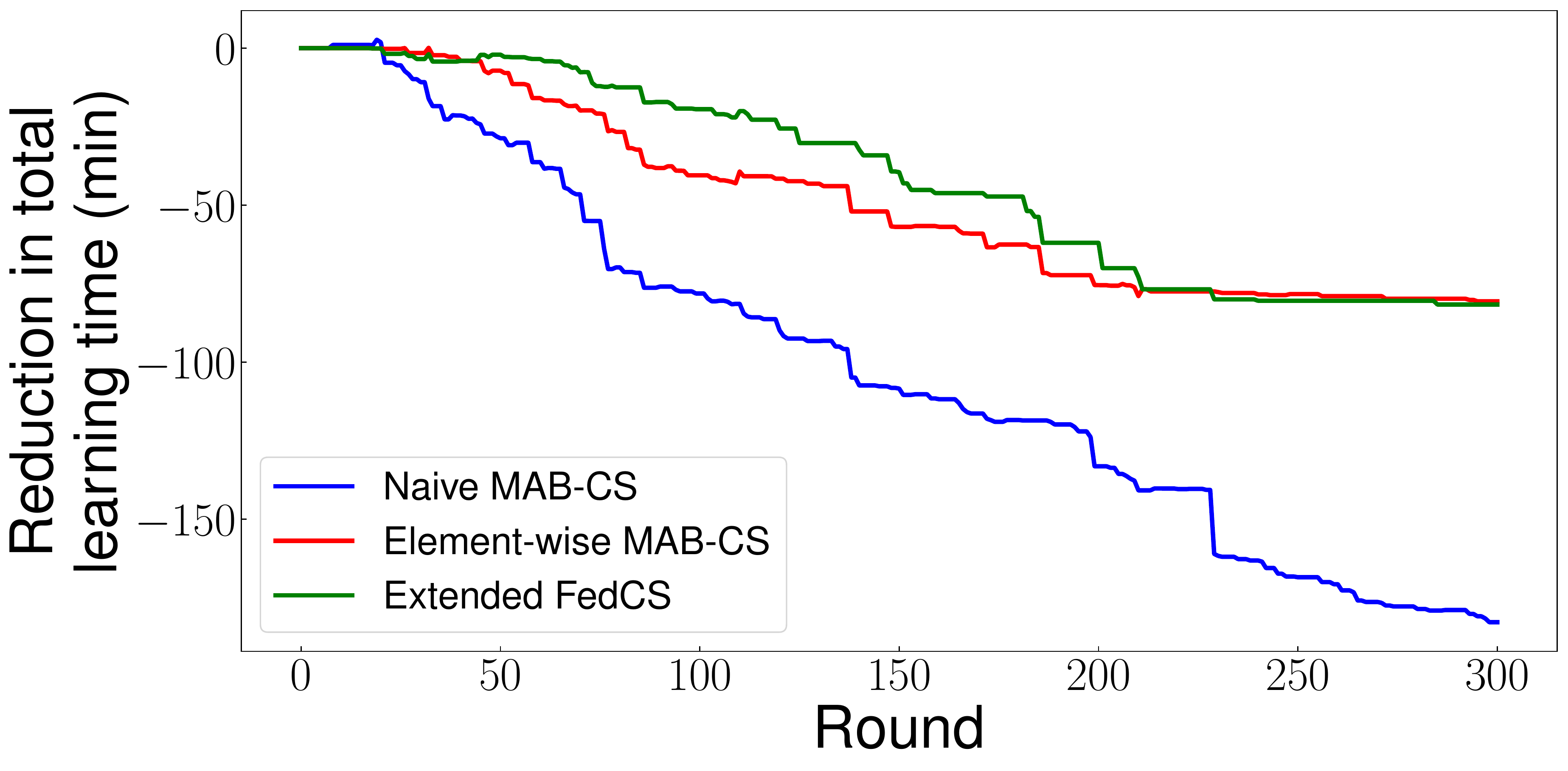}
  }
  \subcaptionbox{$\eta = 1.99 $}{
  \includegraphics[width=0.9\columnwidth]{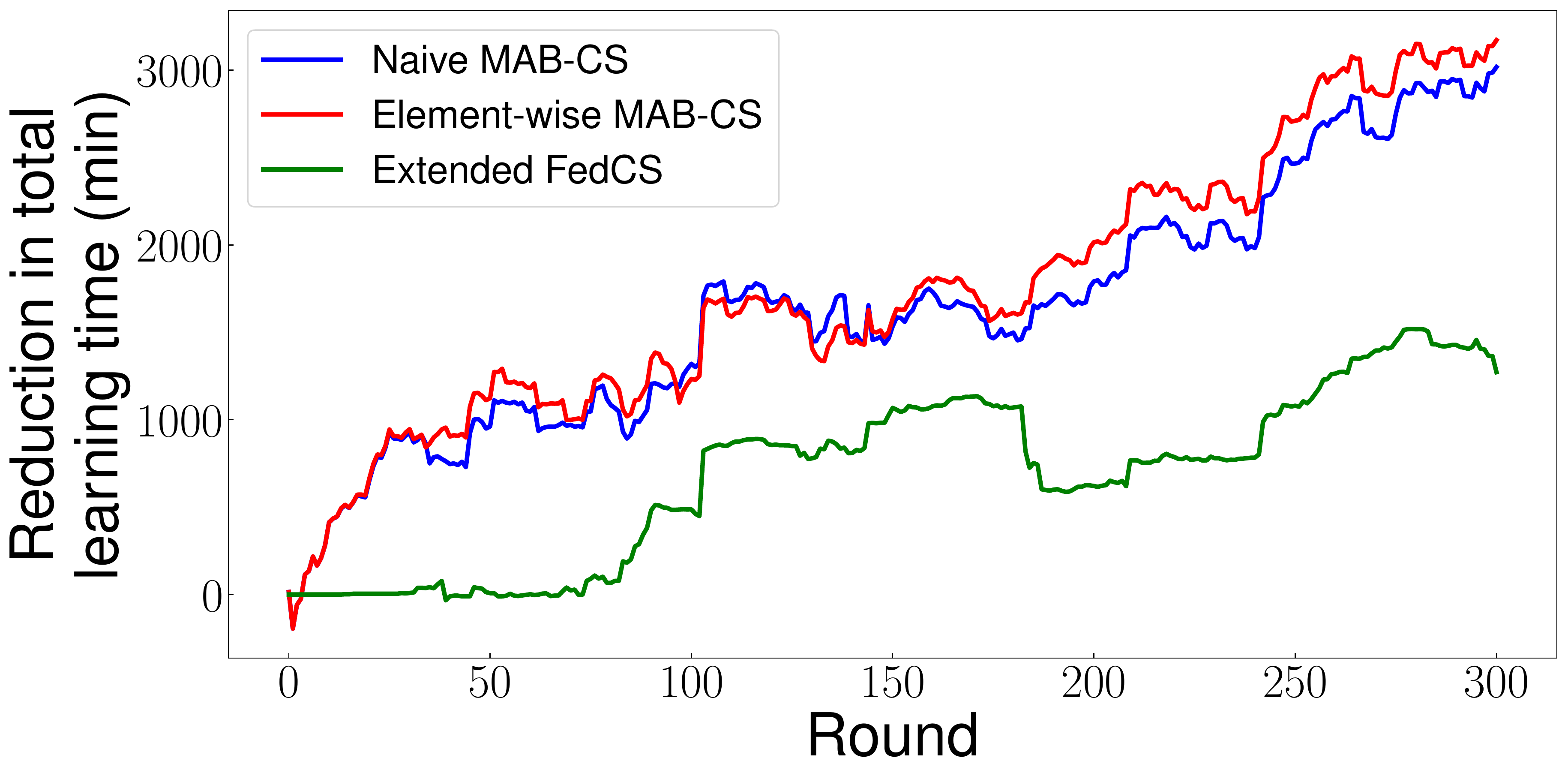}
  }
  \caption{
    Effect of the amount of resource fluctuation on the elapsed time. The vertical axis indicates the difference in elapsed time when compared to FedCS. A larger value on the vertical axis indicates a larger reduction in learning time compared with FedCS.
    }
  \label{fig:Tdiff}
\end{figure}

Further, our model was a standard convolutional neural network and was the same as that employed in \cite{nishio2018client}.
It consisted of six $3\times 3$ convolutional layers (32, 32, 64, 64, 128, and 128 channels, each of which was activated using ReLU and then batch normalized, and every two of which were followed by $2\times 2$ max pooling), followed by three fully connected layers (512 and 192 units activated using ReLU and another 10 units activated using softmax).
This yielded approximately $4.6$ million model parameters ($M$ = 18.3 megabytes in 32-bit float).
When updating models, we selected the following hyperparameters according to the work in \cite{nishio2018client}: $50$ for mini-batch size, $5$ for the number of epochs in each round, $0.25$ for the initial learning rate of SGD updates, and 0.99 for learning-rate decay.

The training dataset was distributed to $K = 100$ clients as follows. First, we randomly determined the number of images owned by each client in a range of 100 to 1,000.
Then, by following the simulation setup used in \cite{zhao2018federated}, we split the training dataset according to the independent identical distributions (IID) setting, where each client randomly samples a specified number of images from the whole training dataset.

\begin{figure}[t]
  \centering
      \includegraphics[width=0.9\columnwidth]{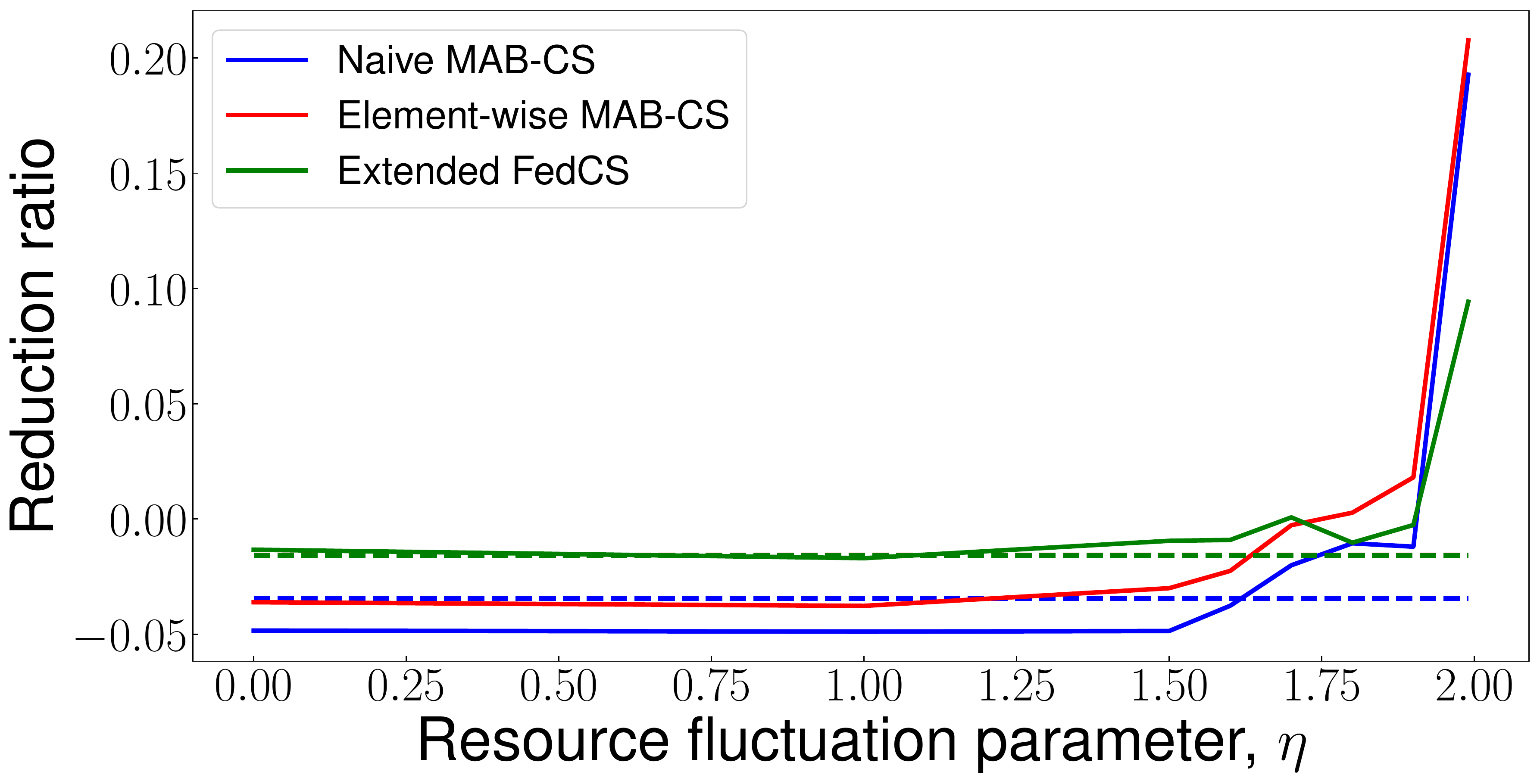}
  \caption{Ratio of reduction in learning time when compared to FedCS with various $\eta$. The dashed lines indicate the reduction ratio when the resources do not fluctuate.}
  \label{fig:ReductionResult}
\end{figure}

\begin{figure}[t]
  \centering
      \includegraphics[width=0.9\columnwidth]{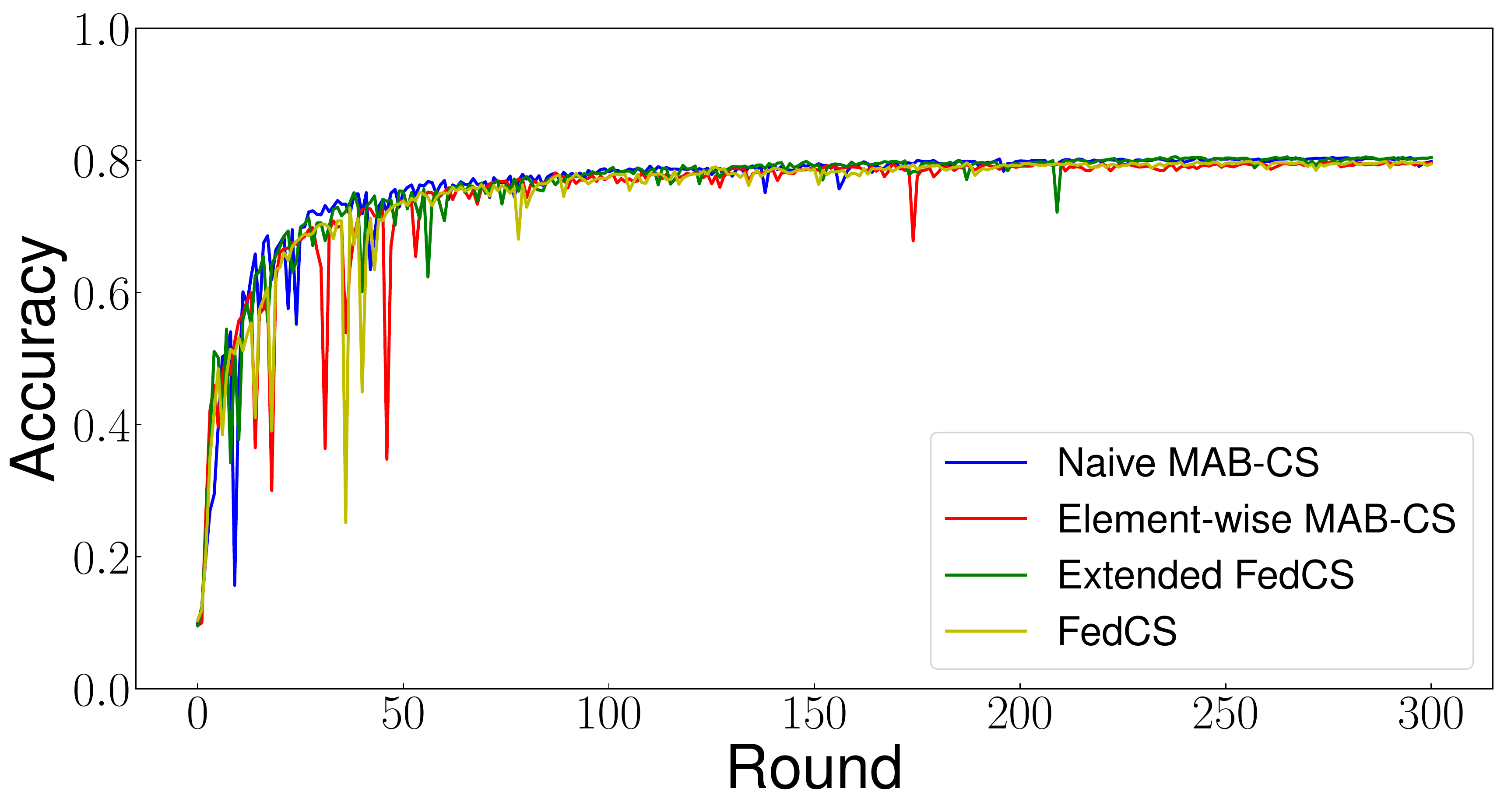}
  \caption{Effect of the client selection algorithms on the prediction accuracy of the trained models when $\eta = 1.5$.}
  \label{fig:acc}
\end{figure}

\subsection{Evaluation Results}
We evaluated three algorithms: \emph{Naive MAB-CS}, \emph{Element-wise MAB-CS}, and \emph{Extended FedCS}.
\emph{Naive MAB-CS} and \emph{Element-wise MAB-CS} are MAB-based selection methods described in Sect.~\ref{sec:proposed}; \emph{Naive MAB-CS} evaluates clients based on Eq. \eqref{eq:evaluation1}; and \emph{Element-wise MAB-CS} evaluates clients using the model update time and model upload time with the UCB amendment described in Eqs. \eqref{eq:tau_UD} and \eqref{eq:tau_UL}.
\emph{Extended FedCS} estimates the expected latency for each client using the moving average of observed latency for model update and uploading, and selects clients based on the estimated latency with the original FedCS algorithm.
This algorithm calculates $T_{\mathrm{inc}}$ based on Eq. \eqref{eq:T_inc}, using the average of the last five selections of model update times and model upload times.
It evaluates clients by using $T_{\mathrm{inc}}$ calculated above as $f(\bm{S},k)$ in Algorithm \ref{alg:scheduling}.
We compared the elapsed times of these three algorithms with that of naive FedCS, which uses the last observed latency as estimated latency of the client.
We also evaluated how the prediction accuracy of models was affected when they were trained using these algorithms.

\begin{figure}[t]
  \centering
  \subcaptionbox{Naive MAB-CS}{
  \includegraphics[width=0.9\columnwidth]{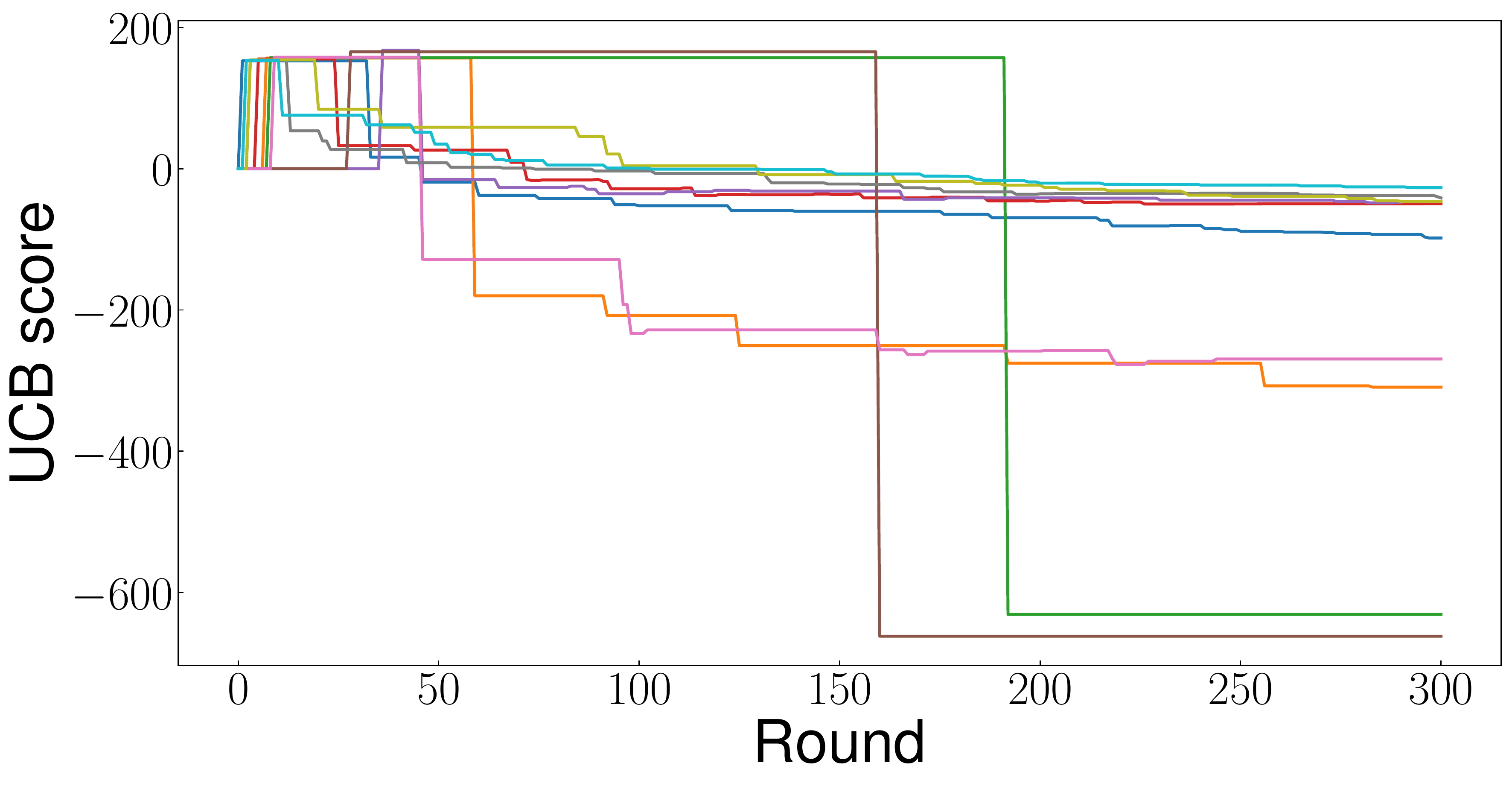}
  }
  \subcaptionbox{Element-wise MAB-CS}{
  \includegraphics[width=0.9\columnwidth]{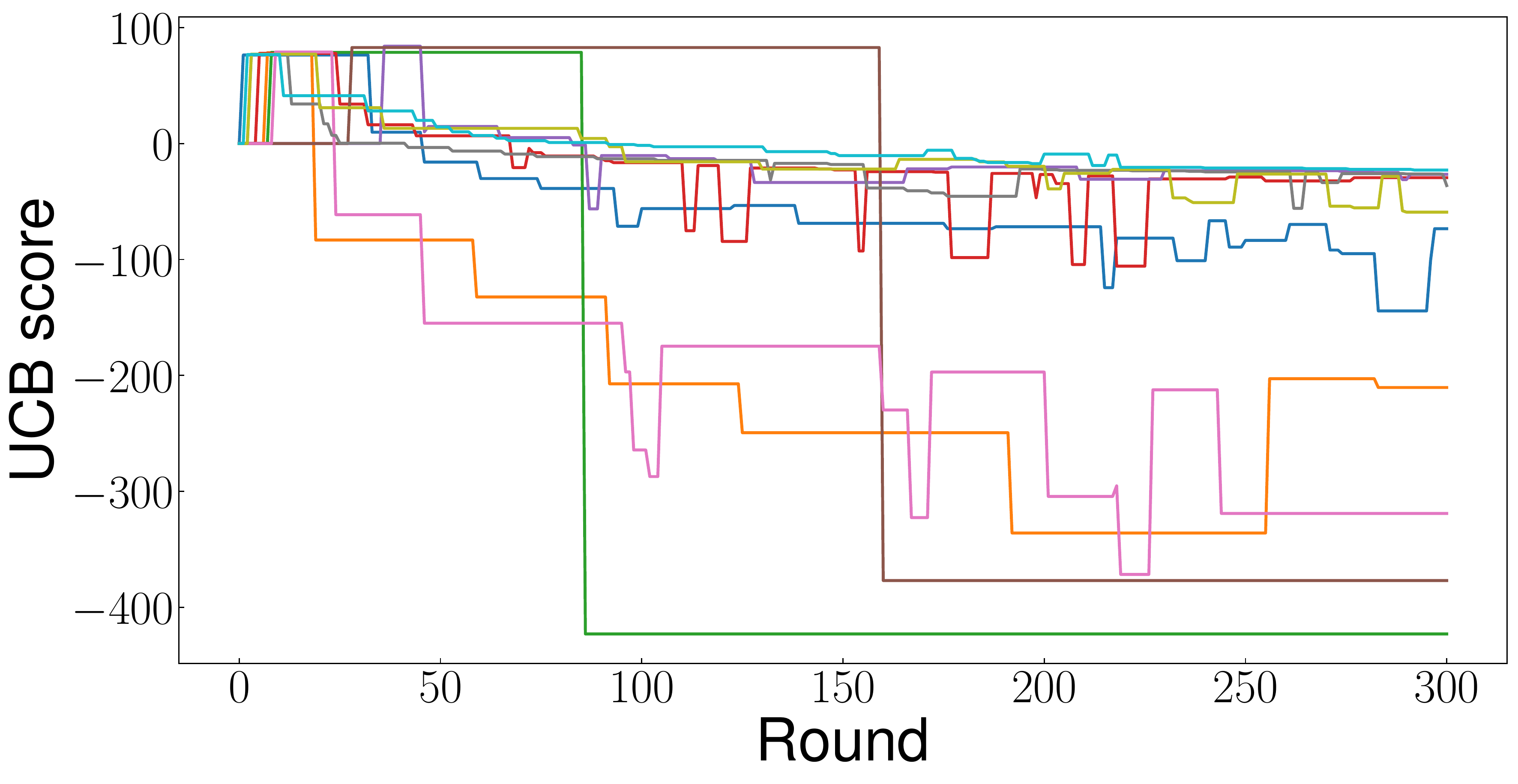}
  }
  \caption{
    Convergence of UCB scores of \emph{Naive MAB-CS} and \emph{Element-wise MAB-CS} when $\eta = 1.5$. The vertical axis indicates the UCB score, $f(\bm{S},k)$. Each graph shows the evaluation values of 10 clients out of 100.
    }
  \label{fig:evaluation}
\end{figure}

Fig.~\ref{fig:Tdiff} illustrates the difference in elapsed time when compared to the FedCS as a function of $\eta$.
The vertical axis indicates the difference in elapsed time,
\begin{align}
  T_{\mathrm{FedCS}}-T_{\mathrm{*}},
\end{align}
where $T_{\mathrm{FedCS}}$ and $T_{\mathrm{*}}$ indicate the elapsed time of FedCS and the illustrated algorithm, respectively.
As shown in Fig.~\ref{fig:Tdiff}(a), when the resources did not fluctuate, the training time was longer than that of FedCS for all three algorithms.
Among the three algorithms, the increase in training time by \emph{Naive MAB-CS} was the largest.
When $\eta$ is relatively large and resources fluctuate widely, the learning times of the three algorithms tend to be shorter than that of FedCS.
In particular, when $\eta=1.99$, the learning times of \emph{Naive MAB-CS} and \emph{Element-wise MAB-CS} were significantly shorter than those of FedCS as shown in Fig.~\ref{fig:Tdiff}(b).
Fig.~\ref{fig:ReductionResult} illustrates the ratio of reduction in learning time with various $\eta$.
The dashed lines show the reduction ratio in a setting with no resource fluctuation.
When the resources fluctuate widely, \emph{Element-wise MAB-CS} exhibits the largest reduction ratio.
The increases in learning time exhibited by \emph{Naive MAB-CS} and \emph{Element-wise MAB-CS} in settings with minor resource fluctuation stem from the fact that exploration, which selects clients that are selected less frequently, is not required in such settings.
When the resources fluctuate widely, we should select clients based on the average values of their resources, rather than the temporary resource information obtained during resource fluctuation; in such settings, the importance of exploration increases.

Fig.~\ref{fig:acc} illustrates how the client selection algorithms affect the prediction accuracy of trained models when $\eta = 1.5$.
All algorithms yielded similar prediction accuracy.
Our proposed algorithms reduced the learning time without affecting the learning accuracy.

Fig.~\ref{fig:evaluation} illustrates the UCB scores of \emph{Naive MAB-CS} and \emph{Element-wise MAB-CS} when $\eta = 1.5$.
The vertical axis indicates the UCB score, $f(\bm{S},k)$.
This result shows that the evaluation values of the clients converged to a certain value.
It shows that MAB-based selection algorithms can obtain certain policies.
In addition, because the two algorithms assigned their largest evaluation values to different clients, it is clear that they follow different client selection policies.

\section{Conclusion}\label{sec:Conclusion}
This paper presented a MAB-based client selection method that balances the trade-off between exploration and exploitation caused in the client selection problem in which the client resources fluctuate and are uncertain until the FL round is conducted.
Our formulation and algorithm design, which fits with FL client selection, enables reduction of the FL latency induced by selecting resource-poor clients.
Subsequently, we conducted simulations by performing realistic ML tasks to demonstrate the effectiveness of our protocol.
Our simulation results revealed that our proposed algorithm achieved shorter learning times than FedCS when resources fluctuated widely.
Future work should consider a setting in which clients' average resource usage will fluctuate during an FL operation.

\section*{Acknowledgment}
This work was supported in part by the KDDI Foundation.

\bibliographystyle{IEEEtran}
\bibliography{GLOBECOMWORKSHOP}

\end{document}